\documentstyle[11pt]{article}

\newlength{\dinwidth}
\newlength{\dinmargin}
\def\slepton{\widetilde \ell}

\def\slr{\slepton_{R}}
\def\sll{\slepton_{L}}
\def\squark{\widetilde q}
\def\msl{m_{\slepton}}
\def\msq{m_{\squark}}
\def\msll{m_{\sll}}
\def\mslr{m_{\slr}}
\def\photino{\widetilde \gamma}
\def\zino{{\widetilde{Z}}}
\def\wino{{\widetilde{W}}}
\def\sfermion{\widetilde f}

\def\gluino{\widetilde g}

\def\sneutrino{\widetilde \nu}

\def\sql{\squark_{L}}

\def\msn{m_{\sneutrino}}
\def\msg{m_{\gluino}}

\def\su{\widetilde{u}}
\def\sd{\widetilde{d}}
\def\sc{\widetilde{c}}
\def\ss{\widetilde{s}}
\def\sul{\su_{L}}
\def\sdl{\sd_{L}}
\def\sur{\su_{R}}
\def\sdr{\sd_{R}}
\def\sclr{\sc_{L,R}}
\def\sslr{\ss_{L,R}}
\def\sulr{\su_{L,R}}
\def\sdlr{\sd_{L,R}}
\def\msul{m_{\sul}}
\def\msur{m_{\sur}}
\def\msdl{m_{\sdl}}
\def\msdr{m_{\sdr}}
\def\msclr{m_{\sclr}}
\def\msulr{m_{\sulr}}
\def\msdlr{m_{\sdlr}}
\def\msslr{m_{\sslr}}
\def\mch{m_{H^{+}}}
\def\mA{m_{A}}
\def\mH{m_{H}}
\def\st{\widetilde{t}}
\def\sb{\widetilde{b}}

\def\sz1{{\widetilde{Z}}_{1}}
\def\szs{{\widetilde{Z}}_{2}}
\def\szt{{\widetilde{Z}}_{3}}
\def\szf{{\widetilde{Z}}_{4}}
\def\swl{{\widetilde{W}}_{1}}
\def\swh{{\widetilde{W}}_{2}}

\def\msz1{m_{\sz1}}
\def\mszs{m_{\szs}}
\def\mszt{m_{\szt}}
\def\mszf{m_{\szf}}
\def\mswl{m_{\swl}}
\def\mswh{m_{\swh}}

\def\gev{{\rm GeV}}

\def\rs{{\sqrt{s}}}
\def\tanbe{\tan\beta}
\def\nle{{\stackrel{<}{\sim}}}
\def\nge{{\stackrel{>}{\sim}}}
\def\goto{\rightarrow}

\def\mt{m_{t}}
\def\mb{m_{b}}
\def\mh{m_{h}}
\def\stl{\st_{1}}
\def\stls{\st^{*}_{1}}
\def\sth{\st_{2}}
\def\sbl{\sb_{1}}
\def\sbh{\sb_{2}}
\def\mstl{m_{\stl}}
\def\msth{m_{\sth}}
\def\msbl{m_{\sbl}}
\def\msbh{m_{\sbh}}
\def\mz{m_{Z}}

\def\tht{\theta_{t}}

\def\tew{\theta_{W}}
\def\cbar{\overline{c}}
\def\sw{\sin^{2}\theta_{W}}
\def\cw{\cos^{2}\theta_{W}}
\def\muo{\mu_{\infty}}
\def\mgo{M_{\infty}}
\def\afo{A_{\infty}}
\def\mfo{m_{\infty}}
\def\mgut{M_{X}}
\def\mtil{\widetilde{m}}
\def\misEt{\slash\hspace{-8pt}E_{T}}

\begin{document}
{}~~~\\
% \vspace{10mm}
\begin{flushright}
ITP-SU-93/05 \\
RUP-93-9
\end{flushright}
\begin{center}
  \begin{Large}
   \begin{bf}
Can Stop be Light Enough ?
 \\
   \end{bf}
  \end{Large}
  \vspace{5mm}
  \begin{large}
    Tadashi Kon
\footnote{
e-mail address : d34477@jpnac.bitnet, mtsk@jpnkektr.bitnet
}
and Toshihiko Nonaka $^{*}$
\footnote{
e-mail address : rra001c@jpnrky00.bitnet
}
\\
  \end{large}
  \vspace{3mm}
Faculty of Engineering, Seikei University, Tokyo 180, Japan \\
$^{*}$Department of Physics, Rikkyo University, Tokyo 171, Japan \\
  \vspace{5mm}
\end{center}
\begin{quotation}
\noindent
\begin{center}
{\bf Abstract}
\end{center}
We examine a possibility
for existence of a light supersymmetric partner of the top quark (stop)
with mass 15 $\sim$ 20 GeV in the
framework of the minimal supergravity GUT model.
Such light stop could explain the
slight excess of the high $p_{T}$ cross section of the $D^{*\pm}$-meson
production in two-photon process at TRISTAN.
We point out that the existence of such stop could change the dominant
decay mode of some particles and could weaken substantially
present experimental bounds on the supersymmetric parameter space.
It seems that there is a finite parameter region allowing existence
of the light stop even if we consider the present experimental data.
Inversely, if the light stop was discovered at TRISTAN,
masses and mixing parameters of
the other SUSY partners as well as masses of the Higgs and
the top will be severely constrained, for example,
$\msg\simeq75\gev$, $\mswl\nle55\gev$, $90\gev\nle\msl\nle130\gev$,
$100\gev\nle\msq\nle150\gev$, $\mh\nle60\gev$ and
$115\gev\nle\mt\nle135\gev$.
Some phenomenological implications on the present and future
experiments are also discussed.
\end{quotation}
\section{\it Introduction}

  Recently, Enomoto et al. in TOPAZ group at TRISTAN have reported
the slight excess of the high $p_{T}$ cross section of the $D^{*\pm}$-meson
production in two-photon process \cite{Enomoto}.
While the disagreement between the measured value and the standard
model prediction is $1.5\sigma$ level, there is a exciting way to
interpret this enhancement.
It is the pair production of the supersymmetric (SUSY) partner of
the top quark (stop) at $e^+e^-$ collision,
$e^+e^-\goto\stl\stls$ \cite{HK,DH,kon}.
Since the stop will decay into the charm-quark plus the lightest
neutralino \cite{HK}, which can be regarded as the lightest
SUSY particle (LSP),
the signature of events will be the charm-quark pair plus large
missing momentum.
This signature would be resemble to the charmed-hadron production
at the two-photon process at $e^+e^-$ colliders.
Enomoto et al. have pointed out that the stop with mass about 15GeV
and the neutralino with mass about 13GeV could well explain the
experimental data.

  It is natural to ask, "Have not already been such light stop
and neutralino excluded by LEP or Tevatron experiments ?" and
"Could such light stop be favored theoretically ?"
In this paper we examine the possibility for existence of the light stop
and the neutralino in the minimal supergravity GUT (MSGUT) scenario
\cite{sugra,RGE} taking into account of the present experimental
bounds on the SUSY parameter space.
Here we consider only the bounds from collider data and do not
concern rather model dependent bounds from the proton decay
experiments and the dark matter searches.

\section{\it Light stop : its theoretical bases}

In the framework of the MSSM \cite{Nilles},
the stop mass matrix in the ($\st_{L}$, $\st_{R}$) basis is expressed by
\begin{equation}\renewcommand{\arraystretch}{1.3}
{\cal M}^{2}_{\st}=\left(
                 \begin{array}{cc}
                   m^{2}_{\st_{L}} & a_{t}m_{t} \\
                   a_{t}m_{t} & m^{2}_{\st_{R}}
                 \end{array}
                \right),
\label{matrix}
\end{equation}
where $\mt$ reads the top mass.
The SUSY mass parameters $m_{\st_{L, R}}$ and $a_{t}$
are parametrized in the following way \cite{susy} :
\begin{eqnarray}
m^{2}_{\st_{L}}&=&{\widetilde{m}}^{2}_{Q_{3}}
  +\mz^{2}\cos{2\beta}\left({\frac{1}{2}}-
{\frac{2}{3}}\sin^{2}\tew\right)+m^{2}_{t}, \label{mstl}\\
m^{2}_{\st_{R}}&=&{\widetilde{m}}^{2}_{U_{3}}
  +{\frac{2}{3}}\mz^{2}\cos{2\beta}\sin^{2}\tew+m^{2}_{t}, \label{mstr}\\
a_{t}&=&A_{t}+\mu\cot\beta,
\end{eqnarray}
where $\tanbe$, $\mu$ and $A_{t}$ denote
the ratio of two Higgs
vacuum expectation values ($=v_{2}/v_{1}$), the SUSY Higgs mass
parameter and
the trilinear coupling constant, respectively.
The soft breaking masses of third generation doublet
${\widetilde{m}}_{Q_{3}}$ and the up-type singlet
${\widetilde{m}}_{U_{3}}$ squarks are related to those
of first (and second) generation squarks as
\begin{eqnarray}
{\widetilde{m}}^{2}_{Q_{3}}&=&{\widetilde{m}}^{2}_{Q_{1}}-I, \label{mq3}\\
{\widetilde{m}}^{2}_{U_{3}}&=&{\widetilde{m}}^{2}_{U_{1}}-2I, \label{mu3}
\end{eqnarray}
where $I$ is a function proportional to the top quark Yukawa
coupling $\alpha_{t}$ and is determined by the renormalization
group equations in the MSGUT.
Throughout of this paper we adopt the notation in Ref.\cite{Hikasa}.

   There are two origins for lightness of the stop compared to
the other squarks and sleptons,
{\romannumeral 1}) smallness of the diagonal soft
masses $m^{2}_{\st_{L}}$ and $m^{2}_{\st_{R}}$ and
{\romannumeral 2}) the left-right stop mixing.
Both effects are originated from the large Yukawa interaction of the
top.
The origin {\romannumeral 1}) can be easily seen from
Eqs.(\ref{mstl})$\sim$(\ref{mu3}).
The diagonal mass parameters $m^{2}_{\st_{L}}$ and $m^{2}_{\st_{R}}$
in Eq.(\ref{matrix}) have possibly small values
owing to the negative large
contributions of $I$ proportional to $\alpha_{t}$ in
Eqs.(\ref{mq3}) and (\ref{mu3}).
It should be noted that this contribution is also important in
the radiative SU(2)$\times$U(1) breaking in the MSGUT.
The Higgs mass squared has similar expression to
Eqs.(\ref{mq3}) and (\ref{mu3}) ;
\begin{equation}
{\widetilde{m}}^{2}_{H_{2}}={\widetilde{m}}^{2}_{L_{1}}-3I,
\label{Higgsmass}
\end{equation}
where ${\widetilde{m}}^{2}_{L_{1}}$ denotes
the soft breaking mass of first generation doublet slepton.
The large contribution of $I$ enables ${\widetilde{m}}^{2}_{H_{2}}$
to become negative at appropriate weak energy scale.
In order to see another origin {\romannumeral 2}) we should diagonalize
the mass matrix Eq.({\ref{matrix}).
The mass eigenvalues are obtained by
\begin{equation}
m^{2}_{\stl\atop\sth}
         ={\frac{1}{2}}\left[ m^{2}_{\st_{L}}+m^{2}_{\st_{R}}
             \mp \left( (m^{2}_{\st_{L}}-m^{2}_{\st_{R}})^{2}
            +(2a_{t}m_{t})^{2}\right)^{1/2}\right].
\label{stopmass}
\end{equation}
and the corresponding mass eigenstates are expressed by
\begin{equation}
\left({\stl\atop\sth}\right)=
\left(
{\st_{L}\,\cos\tht-\st_{R}\,\sin\tht}
\atop
{\st_{L}\,\sin\tht+\st_{R}\,\cos\tht}
\right),
\end{equation}
where $\tht$ denotes the mixing angle of stops :
\begin{eqnarray}
\sin 2\tht=
{\frac{2a_{t}\,m_{t}}
  {\sqrt{(m^{2}_{\st_{L}}-m^{2}_{\st_{R}})^{2}
          +4a^{2}_{t}\,m^{2}_{t}}}}, \\
\cos 2\tht=
{\frac{m^{2}_{\st_{L}}-m^{2}_{\st_{R}}}
  {\sqrt{(m^{2}_{\st_{L}}-m^{2}_{\st_{R}})^{2}
          +4a^{2}_{t}\,m^{2}_{t}}}}.
\label{sintht}
\end{eqnarray}
We see that if SUSY mass parameters and the top
mass are the same order of magnitude,
small $\mstl$ is possible owing to the cancellation
in the expression Eq.~(\ref{stopmass}) \cite{HK,stop}.

After the mass diagonalization
we can obtain the interaction Lagrangian in terms of the
mass eigenstate $\stl$.
We note, in particular, that the stop coupling to
the $Z$-boson ($\stl\stls Z$) depends sensitively on
the mixing angle $\tht$.
More specifically, it is proportional to
\begin{equation}
C_{\stl}\equiv {\frac{2}{3}}\sin^{2}\tew - {\frac{1}{2}}\cos^{2}\tht.
\label{c}
\end{equation}
Note that for a special value of $\tht$$\sim$0.98,
the $Z$-boson coupling completely vanishes \cite{DH}.

\section{\it Present bounds on stop mass}

Before discussion of experimental bounds on the stop mass $\mstl$,
we examine the decay modes of the stop.
In the MSSM, the stop
lighter than the other squarks and gluino
can decay into the various final states :
$\stl$ $\to$ $t\,\sz1$ (a),
$ b\swl$ (b),
$ b\ell\sneutrino $ (c),
$ b\nu\slepton $ (d),
$ bW\sz1 $ (e),
$ bff'\sz1 $ (f),
$ c\sz1$ (g),
where $\sz1$, $\swl$, $\sneutrino$ and $\slepton$, respectively, denote
the lightest neutralino, the lighter chargino, the sneutrino and the
charged slepton.
If we consider the light stop with mass lighter than 20GeV,
the first five decay modes (a) to (e) are kinematically
forbidden due to the model independent
lower mass bounds for respective particles
;  $m_{t}$$\nge$90GeV,
$m_{\swl}$$\nge$45GeV, $m_{\slepton}$$\nge$45GeV and
$m_{\sneutrino}$$\nge$40GeV.
So there left (f) and (g).
Hikasa and Kobayashi \cite{HK} have shown that
the one-loop mode $\stl\to c\sz1$ (g) dominates over the
four-body mode $\stl\to bff'\sz1$ (f).
It is absolutely true in the case considered here, because
the mode (f) is negligible by the kinematical suppression,
$\mstl\sim\msz1 +\mb$.
So we can conclude that such
light stop will decay into
the charm quark jet plus the missing momentum taken away
by the neutralino with  almost 100$\%$ branching ratio.
Note that the width of stop in this case is very small,
i.e., the order of magnitude of eV.

Naively, it will be expected that
Tevatron and/or LEP can set severe
bounds on the stop mass through the processes ;
$gg$ $\to$ $\stl\stl^{*}$ $\to $ $c\cbar\sz1\sz1$
(Tevatron) and/or
$Z$ $\to$ $\stl\stl^{*}$ (LEP).
However, the situation is not so obvious.
Baer et al. \cite{Baer} have performed the analyses of the
experimental data of 4pb$^{-1}$ integrated luminosity
Tevatron running,
and have obtained the results that the stop could easily be escaped
the detection if $m_{\sz1}$ $\nge$ 10GeV.
Such large neutralino mass could make the charm quark jets softer.
Consequently the stop production cross section plotted against
the missing transverse energy becomes smaller than
the present upper bounds,
where they impose cuts on the missing transverse energy.
Moreover, we should point out that LEP cannot exclude the light stop
for appropriate mixing angle $\tht$.
In Fig.1, we show the excluded region in
($\tht$, $\mstl$) plane by LEP in terms of
$\Delta\Gamma_{Z}<35.1$MeV (95\% C.L.) \cite{L3},
where we included the QCD correction in the calculation \cite{DH}.
We find that there is no bound on the stop mass if the
mixing angle $\tht$ is larger than about 0.6.
The origin of such sensitivity of $\Gamma(Z\to\stl\stl^{*})$
is in the fact that the $\stl\stls Z$ coupling is proportional to
$C_{\stl}$ (\ref{c}) \cite{DH}.
TRISTAN have ever settled the lower mass bounds on squarks
$\msq\nge25\gev$ assuming massless photino
in terms of the direct search
$e^{+}e^{-}\goto\squark\squark^{*}$ \cite{TRISTAN}.
Those bounds, however, are invalidated if $\msq-\msz1<8\gev$.
Although we know a bound from direct $\stl$ search at DELPHI reported
by Fisher \cite{Fisher}, we do not concern it here because the
adopted value of mass difference $\mstl-\msz1$ in their analyses
is unknown for us.
Recently Okada \cite{bsg} has investigated
possible bounds on masses of the stop and the neutralino
from the experimental data of the $b\goto s\gamma$ decay.
He has shown that the light stop with mass $\mstl\nle$20GeV
has not been excluded by the data.
After all, we can conclude that there is no bound on the stop mass for
$\msz1\nge$10GeV and $\tht\nge$0.6 if $\mstl-\msz1<8\gev$.

\section{\it Present bounds on gaugino parameters}

In the MSSM, masses and mixing parameters of the gaugino-higgsino
sector are determined by three parameters
$\mu$, $\tanbe$ and $M_{2}$, where $M_2$ denotes the soft breaking
SU(2) gaugino mass.
Some regions in ($\mu$, $\tanbe$, $M_2$) parameter space
have already excluded
by the negative searches for the SUSY particles at
some collider experiments.
First, we concern the experimental data at LEP ;
{\romannumeral 1}) lower bound on the mass of lighter chargino,
$\mswl\nge$45GeV,
{\romannumeral 2}) upper bound on the branching ratio of
the visible neutralino modes of the $Z$, BR$(Z\goto vis.)$ $\equiv$
$\sum_{{i,j}\atop{i=j\neq 1}}\Gamma(Z\goto\zino_{i}\zino_{j})/
\Gamma_{Z}^{\rm tot}$ $<$ $5\times 10^{-5}$ \cite{Lopez}, and
{\romannumeral 3}) upper bound on the invisible width of the $Z$,
$\Gamma(Z\goto\zino_{1}\zino_{1})$ $<$ $16.2$MeV \cite{L3}.
In Fig.2 we show the region excluded by the experimental data
{\romannumeral 1}) $\sim$ {\romannumeral 3}) in ($\mu$, $M_2$)
plane for $\tanbe=$2.
We also plot a contour of $\msz1=13$GeV which can explain the
TRISTAN data as mentioned above.
First we realize that the neutralino with mass $13$GeV can be
allowed in the range
$-160\gev$ $\nle$ $\mu$ $\nle$ $-110\gev$ for $\tanbe=$2.
Note that the contour of $\msz1=13$GeV lies in the excluded region
for $\mu>0$.
If we take larger (smaller) values of $\tanbe$, the allowed
region become narrower (wider).
We find that the allowed region disappears
for $\tanbe\nge 2.3$.
Second we see that $\msz1=13$GeV corresponds to $M_{2}\simeq 22$GeV
in the allowed region and
we can find that this correspondence is
independent on the values of $\tanbe$.
Consequently, we can take $M_{2}=22$GeV as an input value in
the following calculation.
Allowed region in ($\mu$, $\tanbe$) plane fixed by $M_{2}=22$GeV is
shown in Fig.3.
Additional bounds on the ($\mu$, $\tanbe$) parameter space from the
negative search for the neutral Higgs boson at LEP will be
discussed bellow.
It is worth mentioning that the lightest neutralino $\sz1$ is
almost photino $\photino$ in the allowed parameter range
in Fig.3.
In fact, the photino component of the neutralino is larger than
$99$\% in the range.

  Next we should discuss bounds on the gaugino parameters from
the hadron collider experiments.
If we assume {\it the GUT relation},
\begin{equation}
\msg = M_{3} = {\frac{\alpha_{s}}{\alpha}}\sw M_{2}
\label{GUTrel}
\end{equation}
in the MSGUT, the gluino mass $\msg$ bounds from the hadron colliders
could be converted into the bounds on $M_{2}$ \cite{Hidaka}.
The gluino mass bound at CDF taken into account of the cascade decays
$\gluino\goto q\overline{q}\zino_{2,3,4}$ and
$\gluino\goto ud\wino_{1,2}$ \cite{cascade} as well as the direct decay
$\gluino\goto q\overline{q}\zino_{1}$ has reported as \cite{CDF}
\begin{equation}
\msg\nge 95\gev \qquad\quad {\rm (90\% C.L.)}
\label{sgbound}
\end{equation}
for $\mu=-250\gev$ and $\tanbe=2$.
This bound can be easily converted into the bound on $M_2$
by Eq.(\ref{GUTrel}) as $M_{2}\nge 28\gev$, which rejects the
the above fixed value, $M_{2}=22\gev$.
(Note that {\it the GUT relation} (\ref{GUTrel}) depends sensitively
on $\sw$ and $\alpha_{s}$. Here we take $\sw=0.232$ and
$\alpha_{s}=0.113$. )
We must concern, however, a fact that the bound (\ref{sgbound})
is obtained based on the assumption that
$\mstl>\msg$ and the gluino can not decay into the stop.
It is not the case we consider here.
In fact, the gluino can decay into the stop pair,
$\gluino\goto\stl\stls\sz1$, which becomes another seed
for the cascade decay.
In Fig.4 we show the $\msg$ dependence of the branching ratio of
gluino, where we include the mode $\gluino\goto\stl\stls\sz1$
and sum up quark flavors $q, q' = u, d, c, s$.
We take $\tanbe=2$, $\mu=-150\gev$, $\mstl=15\gev$,
$\tht=0.7$, $\mt=135\gev$ and $M_{2}=22\gev$,
and take $\msg$ as a free parameter.
The squark masses are taken as $\msq=2\msg$ (a) and $\msq=3\msg$ (b),
where $\msq$ $\equiv\msulr$ $=\msdlr$ $=\msclr$ $=\msslr$.
An arrow in the figure denotes the $\msg$ value determined by
{\it the GUT relation}.
The branching ratio of the direct decay mode
$\gluino\goto q\overline{q}\zino_{1}$, which is important in the
$\gluino$ search in terms of large $\misEt$
signature, is reduced substantially as
BR$(\gluino\goto q\overline{q}\zino_{1})$ $\nle$ 50\% (15\%),
even for the light gluino with mass $\msg\nge60\gev$ for
$\msq=2\msg$ ($\msq=3\msg$).
Therefore, we should reconsider the UA2 bound $\msg\nge79\gev$
\cite{UA2} obtained under the assumption
BR$(\gluino\goto q\overline{q}\zino_{1})$ $\sim$ 100\%
as well as the CDF bound (\ref{sgbound}).
For the value $\msg =74\gev$ determined by
{\it the GUT relation},
BR$(\gluino\goto q\overline{q}\zino_{1})$ $\sim$ 20\% (4\%)
for $\msq=2\msg$ ($\msq=3\msg$),
which should be compared with
BR$(\gluino\goto q\overline{q}\zino_{1})$ $\sim$ 70\%
obtained when there is no stop mode.
We can find that if we take larger values of $\msq$,
BR$(\gluino\goto q\overline{q}\zino_{1})$ is reduced rapidly.
In this case the Tevatron bound (\ref{sgbound}) would be diminished
significantly.
This is because the width of stop mode
$\Gamma(\gluino\goto\stl\stls\sz1)$ does not depend on $\msq$ but
all the other widths become smaller for larger values of $\msq$.
While all squark masses are independent parameters in the MSSM,
they are determined by small numbers of input parameters in the MSGUT.
Hereafter we adopt {\it the GUT relation} and will reconsider
the Tevatron bound after presenting the results of the MSGUT
analyses.
Note that if we remove {\it the GUT relation}, the gluino can be
heavy with no relation with $M_{2}$ and $\msz1$
and BR$(\gluino\goto q\overline{q}\zino_{1})$ can be small.

\section{\it MSGUT analysis}

   Before presenting our results for the analysis,
we will summarize briefly the calculational scheme in the MSGUT
\cite{Hikasa}.
In this scheme the independent parameters, besides the gauge and Yukawa
couplings, at GUT scale $\mgut$ are the SUSY Higgs mass parameter
$\mu(\mgut)=\muo$ and three soft breaking mass parameters :
the common scalar mass
$\mtil_{\sfermion}^{2}(\mgut)$ $=$ $\mtil_{H_{i}}^{2}(\mgut)$ $=$
$\mfo^{2}$, the common gaugino mass
$M_{3}(\mgut)$ $=$ $M_{2}^{2}(\mgut)$ $=$
$M_{1}^{2}(\mgut)$ $=$ $\mgo$ and
the trilinear coupling
$A_{\tau}(\mgut)$ $=$ $A_{b}(\mgut)$ $=$ $A_{t}(\mgut)$ $=$ $\cdots$
$=$ $\afo$. As usual, we take the Higgs mixing parameter $B$ as
$B(\mgut)=\afo-\mfo$.
All the physical parameters go from $\mgut$ down to low energies
governed by the renormalization group equations (RGE) \cite{RGE}.
In the following we neglect all Yukawa couplings except for the top.
This is not a bad approximation as long as $\tanbe$ is not too large
($\ll\mt /\mb$), which is the case we consider here,
$\tanbe\nle 2.3$.

As for the evolution of the gauge couplings $\alpha_{i}(t)$
and the gaugino masses $M_{i}(t)$, we take the input values
$\alpha^{-1}(\mz)=128.8$ and $\sw=0.232$.
Assuming the SUSY scale is not too different for $\mz$,
we may use the SUSY beta function at all scales above $\mz$
for simplification.
Then one finds $\mgut=2.1\times 10^{16}\gev$, $\alpha_{\infty}^{-1}$
($=\alpha_{3}^{-1}(\mgut)$ $=\alpha_{2}^{-1}(\mgut)$
$=\alpha_{1}^{-1}(\mgut)$) $=24.6$ and
$\alpha_{3}(\mz)=0.113$.
Moreover, the RGE for gaugino masses are easily solved as
\begin{equation}
M_{i}(t)=\alpha_{i}(t){\frac{\mgo}{\alpha_{\infty}}}.
\label{gmass}
\end{equation}

After solving the all other RGE for the physical parameters,
all physics at weak scale $\mz$ are determined by the six
parameters
($\mfo$, $\afo$, $\mgo$, $\mu$, $\tanbe$, $\mt$).
There are, moreover, two conditions imposed on the parameters
to have the correct scale of SU(2)$\times$U(1) breaking.
So we can reduce the number of the independent parameters
to four out of the six.
Here we take the four independent input parameters as
($\mgo$, $\mu$, $\tanbe$, $\mt$).
As we have discussed earlier, furthermore, we can fix one of input value,
$M_{2}=22\gev$, which corresponds to $\mgo=26.7\gev$
for $\sw=0.232$ (see Eq.(\ref{gmass})).
After all, there remain the only three parameters
($\mu$, $\tanbe$, $\mt$).

We seek numerically solutions to give the light stop with mass
$\msz1$ ($=13\gev$) $<$ $\mstl$ $<$ 20GeV varying the three
parameters ($\mu$, $\tanbe$, $\mt$).
The results are shown in Fig.5, which is same parameter space in
Fig.3.
Each hatched area corresponds to the region allowing
$\msz1$ $<$ $\mstl$ $<$ 20GeV for the fixed $\mt$ value.
The upper (lower) line of each area corresponds to
$\mstl=\msz1=13\gev$ ($\mstl=20\gev$).
We also plot the mass $\mh$ contours of the lighter CP-even
neutral Higgs boson as well as the LEP bounds from the data
discussed above.
First we realize that there is rather narrow but finite range
allowing existence of the light stop, if the top was
slightly light too, $\mt\nle 135\gev$.
Second we find that the light stop solutions give inevitably
the light Higgs boson, $\mh\nle 60\gev$.
While we have included the radiative correction in the calculation
of the Higgs mass \cite{higgsmass}, deviations $\delta\mh$ from
the tree level results are not so large, $|\delta\mh|\nle 2\gev$.
The neutral Higgs is standard Higgs like, i.e.,
$\sin(\beta-\alpha)\simeq 1$, where $\alpha$ denotes the Higgs
mixing angle \cite{GH}.
So we should take care the lower mass bound on $\mh$ by
the MSSM Higgs search at LEP,
\begin{equation}
\mh\nge 50\gev \qquad ({\rm 95\% C.L.})
\label{higgsbnd}
\end{equation}
for $\sin(\beta-\alpha)\simeq 1$ \cite{L3}.
Here we must consider the signature of Higgs production at LEP.
We can find that the neutral Higgs will have dominant decay mode
$h\goto \stl\stls$ with almost 100\% branching ratio
as we will discuss bellow.
In this case energies of visible jets from the Higgs would become
softer and it can be smaller than the detection lower cuts.
It is plausible that the bound (\ref{higgsbnd}) would be weakened
although we need a detailed Monte Carlo study
to obtain rigorous bounds from the LEP experiments.
Here we assume a conservative bound $\mh\nge 45\gev$.
Adopting such bound, we can choose three typical parameter sets
(A), (B) and (C), denoted in Fig.5.
The set (A) [(B)] has the largest [smallest] scalar fermion
masses and the smallest [largest] lighter chargino mass.
The set (C) corresponds to the almost center point in the allowed
range.
Input and output values of the parameters of the sets
(A), (B) and (C) are presented in Table 1.
We find that masses and mixing parameters are severely constrained,
for example,
$\mswl$ $\nle$ $55\gev$, $90\gev$ $\nle$ $\msl$ $\nle$ $130\gev$,
$100\gev$ $\nle$ $\msq$ $\nle$ $150\gev$  and
$0.65$ $\nle$ $\tht$ $\nle$ $0.75$.
Note that only the upper limits on $\mswl$ and $\tht$ and
the lower limits on $\msl$ and $\msq$ depend on the tentative
assumption $\mh\nge45\gev$.

\section{\it Phenomenological implications }

  Now we are in position to discuss some consequence of the light stop
scenario in the MSGUT and give strategies to confirm or reject
such possibility in the present and future experiments.
Some numerical results are calculated with the typical parameter sets
(A), (B) and (C).

  The existence of the light stop with mass 15 $\sim$ 20GeV will
alter completely decay patterns of some ordinary and SUSY particles
(sparticles).
First we discuss the top decay \cite{Baer,topdecay}.
In our scenario, the top can decay into final states including the stop;
$t\goto$ $\stl\sz1$, $\stl\szs$ and $\stl\gluino$.
Branching ratios of the top for the typical parameter sets are
presented in Table 2.
We find that the gluino mode $t\goto\stl\gluino$ has
40$\sim$50\% branching ratio and dominates over the
standard mode $t\goto bW^{+}$ in almost whole allowed parameter region.
Strategies for the top search at Tevatron
would be forced to change because the leptonic branching ratios of
the top would be reduced by the dominance of the stop-gluino mode.

   Decay patterns of the Higgs particles will be changed too.
The lighter CP-even neutral Higgs decays dominantly into the stop
pair, $h\goto\stl\stls$, owing to the large Yukawa coupling of the top.
In rough estimation, we obtain
\begin{equation}
{\rm BR}(h\goto\stl\stls) \simeq
{\frac{1}{1+{\frac{3\mb^{2}\mh^{2}}{2\mt^{4}}}}} \simeq 1 .
\end{equation}
This fact would change the experimental methods of the Higgs
searches at the present and future collider experiments.
More detail analyses of the charged \cite{higgsdecay} and
neutral Higgs bosons are presented separately.

  Now we discuss briefly the light stop impact on the sparticle
decays.
The lightest charged sparticle except for the stop
is the lighter chargino $\swl$.
The two body stop mode $\swl\goto b\stl$ would dominate over the
conventional three body mode $\swl\goto f\overline{f'}\sz1$.
As a consequence, it would be difficult to use the leptonic
signature in the chargino search at $e^+e^-$ and hadron colliders.
Since the chargino $\swl$, the neutral Higgs $h$ and
the gluino $\gluino$, whose dominant decay modes are respectively
$\swl\goto b\stl$, $h\goto \stl\stls$ and $\gluino\goto \stl\stls\sz1$,
are copiously produced in the other sparticle decay,
many stops would be expected in the final states of the
sparticle production. For example,
$\sll$ $\goto\nu\swl$ $\goto\nu(b\stl)$,
$\squark_{L,R}$ $\goto q\gluino$ $\goto q(\stl\stls\sz1)$,
$\sql$ $\goto q'\swl$ $\goto q'(b\stl)$ and
$\zino_{i(i\neq 1)}$ $\goto\sz1 h$ $\goto\sz1(\stl\stls)$.
Note that the dominant decay modes of the right-handed sleptons
would be unchanged, i.e.,
BR($\slr\goto\ell\sz1$) $\simeq$ 100\%.

  Needless to say, all experimental groups, AMY, TOPAZ and VENUS,
at TRISTAN should perform a detail data analyses to confirm or
reject the exciting scenario.
Furthermore, we can see that
the stop and its relatively light accompaniments,
the gluino $\gluino$, light neutralinos $\zino_{1,2}$, and
neutral Higgs $h$, should be visible at LEP, SLC, HERA and Tevatron.
   Especially, LEP could search whole allowed region presented in
Figs.1 and 5 in terms of the width of $Z$-boson.
First we find from Table 1,
the stop mixing angle $\tht$ is severely
limited as 0.65 $\nle$ $\tht$ $\nle$ 0.75 in the allowed
range in Fig.5.
It should be noted that the upper bound depends on the tentative
assumption $\mh\nge45\gev$. The larger (smaller) lower limit on $\mh$
gives slightly narrower (wider) range of $\tht$.
On the other hand, the lower bound on $\tht$ is determined by
the established limit $\mswl\nge45\gev$.
It is interesting that $\tht\simeq0.7$ is not {\it input}
but {\it output} of the MSGUT calculation.
The allowed values of $\tht$ with $\mstl =$ 15$\sim$20GeV
are very close to present experimental limit from
$\Delta\Gamma_{Z}$ measurement as depicted in Fig.1.
We find that the stop contribution $\Gamma(Z\goto\stl\stls)$ to
$\Delta\Gamma_{Z}$ is larger than about 15MeV for
$\tht\nle$0.75 and $\mstl\nle$20GeV.
Second, the whole allowed region in Fig.5 can be explored by the
precise measurement of BR($Z\goto vis.$).
In fact, the smallest value of the neutralino contribution
$\sum_{{i,j}\atop{i=j\neq 1}}\Gamma(Z\goto\zino_{i}\zino_{j})/
\Gamma_{Z}^{\rm tot}$ to BR($Z\goto vis.$) is
$1.1\times 10^{-5}$ ($1.8\times 10^{-5}$)
for $\mh>45\gev$ ($\mh>50\gev$).
Of course, the Higgs $h$ search at LEP with the stop signature
$h\goto\stl\stls$ is very important to set further constraint on
the allowed region.
   Clearly, the lighter chargino, $\mswl\nle 55\gev$, would be visible
at LEP{\uppercase\expandafter{\romannumeral 2}}.
Furthermore, there is a possibility that some sleptons
will be discovered at $\rs$ $\nle$ 200GeV.

    As mentioned before, Tevatron will play a crucial role
in confirming or rejecting the light stop scenario in the MSGUT
with {\it the GUT relation}.
In this case the existence of relatively light gluino, $\msg\simeq$ 75GeV,
with substantially large decay fraction $\gluino\goto\stl\stls\sz1$ is
one of definite prediction.
Values of branching ratios of the gluino for the typical parameter sets
(A), (B) and (C) are tabulated in Table 3.
The branching ratio of direct decay mode
BR($\gluino\goto q\overline{q}\sz1$) $=$ $25\sim 50$\% is expected
in the allowed range.
These values are rather large compared to those in Fig.4(b).
This is originated from the fact that allowed mass values of squarks
except for the heavier stop $\sth$ are relatively small
$\msq\nle150\gev$.
Those squarks could be within reach of Tevatron.
Signatures of those squarks, however, would be unusual because of
their cascade decays such as
$\squark_{L,R}$ $\goto q\gluino$ $\goto q(\stl\stls\sz1)$ and
$\sql$ $\goto q'\swl$ $\goto q'(b\stl)$.
Anyway, there is a large possibility that some regions could have
been excluded by Tevatron data.
Detail Monte Carlo studies including the stop mode
$\gluino\goto\stl\stls\sz1$ are required at any cost.
At least we expect that the region near the point corresponds to
the set (B) could have been excluded because
BR$(\gluino\goto q\overline{q}\zino_{1})$ would be large enough
to make the \ $\misEt$ signal events larger than its experimental
upper limits.
Note again that the light stop and neutralino can survive even
after the negative search for the gluino and squarks at Tevatron
if we remove {\it the GUT relation} Eq.(\ref{GUTrel}).
Removal of {\it the GUT relation} corresponds to the change
of boundary conditions on the soft gaugino masses at the unification
scale $M_{X}$.
Owing to this change the RGE solution for the stop mass is modified
and in turn we will get different allowed parameter region
to Fig.5.
The analyses based on such models will be presented elsewhere.

  The $ep$ collider HERA could search the light stop through
its pair production process
$ep\goto e\stl\stls X$ via boson-gluon fusion \cite{stopbg}.
The total cross section of the process is larger than about
$10$pb for $\mstl\nle$20GeV, which is independent on the
mixing angle $\tht$. That is, $\sigma\nge 10$pb is expected
for all parameters with $\mstl\nle$20GeV in the allowed range
in Figs.1 and 5.
Detail analyses with Monte Carlo studies including
possible dominant background process
$ep\goto ec\cbar X$ can be found in Ref.\cite{sthera}.
The process $ep\goto e\stl\stls X$ is useful to confirm or reject
the very light stop, but not to effective in determining model
parameters in the MSSM such as the other sparticle masses as well as
$\tht$.
We have calculated another process
\begin{equation}
ep\goto b\stls\sneutrino X,
\end{equation}
which is expected to be useful in the latter purpose because the
cross section depends on $\mswl$, $\msn$, $\tht$ and
the mixing angles of $\swl$.
The total cross section is obtained as
$1.4\times 10^{-3}$pb for the set (A) and
$2.0\times 10^{-2}$pb for the set (B).
Although these values are rather small and the high luminosity
would be needed to see, sensitivity on the SUSY parameters is
rather high.

Here a comment on the $R$-parity breaking couplings of the stop
may be in order.
We have calculated not only the $R$-conserving process
$ep\goto e\stl\stls X$ but also the $R$-breaking process
$ep\goto (\stl) \goto eq X$ in Refs.\cite{sthera,stoprb}.
Latter process could have a distinctive signature, i.e.,
a sharp peak in the Bjorken parameter $x$ distribution.
In our scenario of a very light stop, however,
the strength of the $R$-parity (and lepton number) breaking
coupling of the stop defined in the superpotential \cite{RB} ;
\begin{equation}
W = \lambda'_{131}L_{1}Q_{3}D_{1}
\label{rbc}
\end{equation}
can not be large enough for possible detection at HERA.
This is because the coupling Eq.(\ref{rbc}) with
$\lambda'_{131}$ $>$ $10^{-4}$ make the $R$-breaking stop decay mode
$\stl\goto ed$ dominant and this fact contradicts
BR($\stl\goto c\sz1$) $\simeq$ 100\% expected at TRISTAN.

Polarized initial electron beams at SLC and at any linear $e^+e^-$
colliders will be efficient to reveal the nature of left-right
mixing in the stop sector, in other words, to measure the mixing angle
of stop $\tht$.
In Fig.6 we show the $\rs$ dependence of the left-right asymmetry ;
\begin{equation}
A_{LR}\equiv{\frac{\sigma(e_{L})-\sigma(e_{R})}
                  {\sigma(e_{L})+\sigma(e_{R})}},
\label{asym}
\end{equation}
where $\sigma(e_{L,R})$ $\equiv$
$\sigma(e^{+}e^{-}_{L,R}\goto\stl\stls)$, which are obtained by
\begin{eqnarray}
&&\sigma(e^{+}e^{-}_{L\atop R}\goto\stl\stls) \\
&&={\frac{\pi\alpha^2}{s}}\beta_{\stl}^{3}
\left[{\frac{4}{9}}+{\frac{2}{3}}C_{\stl}(v_{e}\pm a_{e})
{\rm Re}\left({\frac{s}{D_{Z}}}\right)+
{\frac{1}{4}}C_{\stl}^{2}(v_{e}\pm a_{e})^{2}
\left|{\frac{s}{D_{Z}}}\right|^{2}\right](1+\delta_{QCD}),
\end{eqnarray}
where $\beta_{\stl}$ $\equiv$ $\sqrt{1-4\mstl^{2}/s}$,
$D_{Z}$ $\equiv$ $s-\mz^{2}+i\mz\Gamma_{Z}$,
$v_{e}$ $\equiv$ $(-{\frac{1}{2}}+2\sw)/(\sw\cw)$ and
$a_{e}$ $\equiv$ $-1/(2\sw\cw)$.
In the asymmetric combination in Eq.(\ref{asym}) the photon
contribution is cancelled out and $A_{LR}$ is proportional to
$C_{\stl}$ Eq.(\ref{c}).
This is the reason for sensitive dependence on $\tht$ of $A_{LR}$
in Fig.6.
Another important property is that $A_{LR}$ is independent on the mass
of stop $\mstl$ as well as on the QCD correction $\delta_{QCD}$
since $\beta_{\stl}^{3}$ and $1+\delta_{QCD}$ disappeared in the
fractional combination of $\sigma$ in $A_{LR}$.
Therefore, this method for measuring $\tht$ will be applicable for
the stop with any mass satisfying $\mstl<\rs/2$.

\section{\it Conclusion}

  We have investigated the possibility for existence of the light stop
$\mstl=15\sim20\gev$ and the neutralino $\msz1\simeq13\gev$
in the MSGUT scenario taking into account of the present
experimental bounds on the SUSY parameter space.
We have pointed out that
the existence of such stop could change the dominant
decay mode of some particles.
For example, the stop modes $\gluino\goto\stl\stls\sz1$ and
$h\goto\stl\stls$ could dominate over respectively
the conventional modes $\gluino\goto q{\overline{q}}\sz1$ and
$h\goto b\overline{b}$ even for relatively light gluino and Higgs.
As a consequence, present experimental bounds on the
SUSY parameter space could be weakened considerably.
It seems that there is a finite parameter region allowing existence
of such light stop even if we consider the present experimental data.
Inversely, it has been found that,
if such light stop was discovered at TRISTAN,
masses and mixing parameters of
the other SUSY partners as well as masses of the Higgs and
the top will be severely constrained, for example,
$\msg\simeq75\gev$, $\mswl\nle55\gev$, $90\gev\nle\msl\nle130\gev$,
$100\gev\nle\msq\nle150\gev$, $0.65\nle\tht\nle0.75$,
$\mh\nle60\gev$ and $115\gev\nle\mt\nle135\gev$.
Actually, the light stop and its relatively light accompaniments,
the gluino $\gluino$, the light neutralinos $\zino_{1,2}$, and
the neutral Higgs $h$, should be visible near future
at LEP, HERA and Tevatron.
In fact, LEP and HERA could explore the whole allowed parameter region
in terms of the precise measurement of the width of $Z$ and
by means of searching for the process
$ep\goto e\stl\stls X$, respectively.
There exists, moreover, special interest on Tevatron experiment, i.e.,
Tevatron could either discover or exclude the light gluino
$\msg\simeq 75\gev$.
If we could not discover such gluino,
this fact would indicate invalidity of {\it the GUT relation},
in other words, the assumption of the common gaugino mass at $M_{X}$.
Experimental sign for violation of {\it the GUT relation}
may be important to select a specific model out of a great number of
string model candidates.

 Recently, Altarelli et al. \cite{Altarelli} have shown that
the light stop $\mstl\nle50\gev$ and light chargino
$\mswl\nle60\gev$ could well explain the precision data at LEP.
Their result seems to support our light stop scenario.
In this paper
we have exemplified that if we discover the light stop we will
be able to constrain severely all the SUSY parameters
at the unification scale.
We can conclude that, therefore, the discovery of the stop will bring
us a great physical impact.
Not only it will be the first signature of the top flavor and
the supersymmetry but also it could shed a light on the physics at
the unification scale.

\vskip 20pt
\begin{flushleft}
{\Large{\bf Acknowledgement}}
\end{flushleft}
One of the authors (T.K.) would like to thank Drs. R. Enomoto
and M. Pieri for valuable information.

\vskip 20pt

%\begin{flushleft}
%{\Large{\bf Appendix}}
%\end{flushleft}

\vskip 20pt
%\vfill\eject

\vfill\eject

\baselineskip = 18pt plus 1pt
\noindent{\Large{\bf Figure Captions}} \\
\medskip
\nobreak
{\bf Figure 1:} \ \
Excluded region in
($\tht$, $\mstl$) plane by LEP with
$\Delta\Gamma_{Z}<35.1$MeV. \label{fig1}
\\
\medskip
{\bf Figure 2:} \ \
Contour of $\msz1=13$GeV in ($\mu$, $M_2$) plane for $\tanbe=$2.
Excluded region by LEP is also depicted. \label{fig2}
 \\
\medskip
{\bf Figure 3:} \ \
Allowed region in ($\mu$, $\tanbe$) plane for $M_{2}=22$GeV.
Dashed line and dotted line respectively correspond to
contour of BR($Z\goto vis$) $=$ $5\times 10^{-5}$ and that of
$\mswl=45\gev$, respectively. \label{fig3}
 \\
\medskip
{\bf Figure 4:} \ \
$\msg$ dependence of branching ratios of gluino.
Sum over quark flavors $q,q' = u, d, c, s$ are taken.
Input parameters are $\tanbe=2$, $\mu=-150\gev$, $\mstl=15\gev$,
$\tht=0.7$, $\mt=135\gev$, $M_{2}=22\gev$ and $\msq=2\msg$ (a)
and $\msq=3\msg$ (b).
An arrow in the figure denotes the $\msg$ value determined by
{\it the GUT relation}. \label{fig4}
 \\
\medskip
{\bf Figure 5:} \ \
Stop mass contours in ($\mu$, $\tanbe$) plane for fixed $\mt$.
Each hatched area corresponds to the region
$\msz1$ $<$ $\mstl$ $<$ 20GeV.
The upper (lower) line of each area corresponds to
$\mstl=\msz1$ ($\mstl=20\gev$).
We also plot the mass $\mh$ contours as well as the LEP bounds.
Points denoted by A, B and C are correspond to
typical parameter sets in the text. \label{fig5}
 \\
\medskip
{\bf Figure 6:} \ \
Total energy $\rs$ dependence of left-right asymmetry
for the stop production at $e^+e^-$ colliders.
For comparison we also plot $A_{LR}$ for the up-type quark
production. \label{fig6}
 \\
\vfill\eject

\begin{center}
\begin{tabular}{|c|rrr|}
\multicolumn{4}{c}{{\bf Table 1}: Typical parameter sets}\\
\hline
masses in GeV  &     A    &     B    &     C     \\
\hline
$M_2$   & $22$     & $22$     &  $22$     \\
$\tanbe$& $2.09$   & $1.83$   &  $2.0$    \\
$\mu$   & $-151$   & $-94.5$  &  $-135$   \\
$\mt$   & $135$    & $117$    &  $130$    \\
\hline
\hline
$\mgo$  & $26.7$   & $ 26.7$  &  $ 26.7$  \\
$\mfo$  & $125.1$  & $ 83.6$  &  $111.6$  \\
$\afo$  & $332.2$  & $206.8$  &  $293.3$  \\
$\muo$  & $-132.8$ & $-78.4$  &  $-116.7$ \\
\hline
$\mstl$ & $14.6$   & $15.4$   &  $15.5$   \\
$\msth$ & $201.9$  & $183.0$  &  $196.4$  \\
$\tht$  & $0.66$   & $0.72$   &  $0.68$   \\
$\msbl$ & $107.0$  & $93.0$   &  $102.0$  \\
$\msbh$ & $142.9$  & $108.5$  &  $131.2$  \\
$\msul$ & $136.0$  & $100.4$  &  $124.0$  \\
$\msur$ & $138.5$  & $103.1$  &  $126.6$  \\
$\msdl$ & $150.0$  & $116.3$  &  $138.6$  \\
$\msdr$ & $142.7$  & $107.9$  &  $131.0$  \\
$\msll$ & $132.0$  & $92.5$   &  $119.0$  \\
$\mslr$ & $130.3$  & $90.2$   &  $117.2$  \\
$\msn$  & $115.8$  & $71.5$   &  $101.7$  \\
$\mh$   & $55.5$   & $44.5$   &  $52.6$   \\
$\mA$   & $211.0$  & $135.0$  &  $189.0$  \\
$\mH$   & $224.8$  & $159.1$  &  $205.1$  \\
$\mch$  & $225.6$  & $156.9$  &  $205.2$  \\
$\alpha$& $-0.56$  & $-0.69$  &  $-0.59$  \\
$\msz1$ & $13.1$   & $13.3$   &  $13.2$   \\
$\mszs$ & $48.2$   & $54.7$   &  $50.5$   \\
$\mszt$ & $159.0$  & $107.6$  &  $143.6$  \\
$\mszf$ & $187.3$  & $142.6$  &  $174.2$  \\
$\mswl$ & $45.0$   & $53.6$   &  $47.3$   \\
$\mswh$ & $184.5$  & $138.9$  &  $171.1$  \\
$\msg$  & $74.4$   & $74.4$   &  $74.4$   \\
\hline
\end{tabular}
\label{table1}
\end{center}

\vfill\eject

\begin{center}
\begin{tabular}{|rl|ccc|}
\multicolumn{5}{c}{{\bf Table 2}: Branching ratios of top}\\
\hline
         &              &     A    &     B    &     C      \\
\hline
$t\goto$ & $\stl\sz1$   & $0.054$  & $0.093$  &  $0.063$   \\
$      $ & $\stl\szs$   & $0.049$  & $0.130$  &  $0.056$   \\
$      $ & $\stl\gluino$& $0.516$  & $0.403$  &  $0.493$   \\
$      $ & $bW^{+}$     & $0.381$  & $0.374$  &  $0.388$   \\
\hline
\end{tabular}
\label{table2}
\end{center}

\begin{center}
\begin{tabular}{|rl|ccc|}
\multicolumn{5}{c}{{\bf Table 3}: Branching ratios of gluino}\\
\hline
               &                     &     A    &     B    &     C    \\
\hline
$\gluino\goto$ & $q\overline{q}\sz1$ & $0.248$  & $0.497$  &  $0.342$ \\
$            $ & $q\overline{q}\szs$ & $0.022$  & $0.011$  &  $0.021$ \\
$            $ & $q\overline{q'}\swl$& $0.087$  & $0.044$  &  $0.087$ \\
$            $ & $\stl\stls\sz1$     & $0.643$  & $0.448$  &  $0.550$ \\
\hline
\end{tabular}
\label{table3}
\end{center}
\end{document}